\documentclass[11pt]{article}
\usepackage{amssymb}
\usepackage{epsfig}
\usepackage{graphicx}
\usepackage[nomarkers]{endfloat}
\usepackage{pstricks}

\newcommand{\BE}{\begin{equation}}
\newcommand{\EE}{\end{equation}}
\begin{document}
\begin{titlepage}

\vspace*{1mm}
\begin{center}

{\LARGE{\bf Emergent gravity and ether-drift experiments}}

\vspace*{14mm} {\Large  M. Consoli and L. Pappalardo}
\vspace*{4mm}\\
{\large
Istituto Nazionale di Fisica Nucleare, Sezione di Catania \\
Dipartimento di Fisica e Astronomia dell' Universit\`a di Catania \\
Via Santa Sofia 64, 95123 Catania, Italy \\ }
\end{center}
\begin{center}
{\bf Abstract}
\end{center}

According to several authors, gravity might be a long-wavelength
phenomenon emerging in some `hydrodynamic limit' from the same
physical, flat-space vacuum viewed as a form of superfluid medium.
In this framework, light might propagate in an effective acoustic
geometry and exhibit a tiny anisotropy that could be measurable in
the present ether-drift experiments. By accepting this view of the
vacuum, one should also consider the possibility of sizeable random
fluctuations of the signal that reflect the stochastic nature of the
underlying `quantum ether' and could be erroneously interpreted as
instrumental noise. To test the present interpretation, we have
extracted the mean amplitude of the signal from various experiments
with different systematics, operating both at room temperature and
in the cryogenic regime. They all give the same consistent value
$\langle A \rangle ={\cal O}(10^{-15})$ which is precisely the
magnitude expected in an emergent-gravity approach, for an apparatus
placed on the Earth's surface. Since physical implications could be
substantial, it would be important to obtain more direct checks from
the instantaneous raw data and, possibly, with new experimental
set-ups operating in gravity-free environments.

\end{titlepage}

\section{Introduction}

According to the generally accepted view, gravitational phenomena
are described through the introduction of a non-trivial local metric
field $g_{\mu\nu}(x)$ which is interpreted as a fundamental
modification of the flat space-time of Special Relativity. By {\it
fundamental}, one means that, in principle, deviations from flat
space might also occur at arbitrarily small scales, e.g. down to the
Planck length.

However, it is an experimental fact that many physical systems
(moving fluids, condensed matter systems with a refractive index,
Bose-Einstein condensates,...) for which, at a fundamental level,
space-time is exactly flat, are nevertheless described by an
effective curved metric in their hydrodynamic limit, i.e. at length
scales that are much larger than the size of their elementary
constituents. For this reason, one could try to explore the
alternative point of view where the space-time curvature observed in
gravitational field emerges \cite{barcelo1,barcelo2} in a similar
way from hydrodynamic distortions of the same physical, flat-space
vacuum viewed as a form of superfluid ether \cite{volo}
('emergent-gravity' approach).

In this different perspective, local re-scalings of the basic
space-time units could represent the crucial ingredient to generate
an effective non-trivial curvature, see e.g. \cite{feybook,dicke1},
in view of the substantial equivalence with the standard
interpretation  : "It is possible, on the one hand, to postulate
that the velocity of light is a universal constant, to define {\it
natural} clocks and measuring rods as the standards by which space
and time are to be judged and then to discover from measurement that
space-time is {\it really} non-Euclidean. Alternatively, one can
{\it define} space as Euclidean and time as the same everywhere, and
discover (from exactly the same measurements) how the velocity of
light and natural clocks, rods and particle inertias {\it really}
behave in the neighborhood of large masses" \cite{atkinson}.

Although one does not expect to reproduce exactly the same features
of classical General Relativity, still there is some value in
exploring this possibility. In fact, beyond the simple level of an
analogy, there might be a deeper significance if the properties of
the underlying ether, that are required by the observed metric
structure, could be matched with those of the physical vacuum of
electroweak and strong interactions. In this case, the so called
vacuum condensates, that play a crucial role for fundamental
phenomena such as mass generation and quark confinement, could also
represent a bridge between gravity and particle physics.

For a definite realization of this idea, one could then start by
representing the physical quantum vacuum as a Bose condensate of
elementary quanta and look for vacuum excitations that, on a coarse
grained scale, resemble the Newtonian potential. In this case, it is
relatively easy \cite{ultraweak} to match the weak-field limit of
classical General Relativity or of some of its possible variants.
The idea that Bose condensates can provide various forms of
gravitational dynamics is not new (see e.g.
\cite{bosegravity,sindoni} and references quoted therein) and it is
conceivable that the analogy could also be extended to higher
orders. Therefore, being faced with two completely different
interpretations of the same metric structure, one might ask: could
this basic conceptual difference have phenomenological implications
? The main point of our paper is that, in principle, a
phenomenological difference might be associated with a small
anisotropy of the velocity of light in the vacuum and that this tiny
effect is within of reach of the present generation of precise
ether-drift experiments.

After this general introduction, the plan of the paper is as
follows. In Sect. 2, we shall illustrate this theoretical framework
in a class of emergent-gravity scenarios, those in which the
effective curvature is indeed induced by a re-definition of the
basic space-time units and by a non-trivial vacuum refractive index.
Although, in many respects, this perspective is equivalent to the
conventional point of view, there is however a notable difference:
the speed of light in the vacuum might not coincide with the basic
parameter $c$ entering Lorentz transformations. On a general ground,
this opens the possibility of a non-zero light anisotropy whose
typical fractional magnitude, for an apparatus placed on the Earth's
surface, can be estimated to be ${\cal O}(10^{-15})$.

Checking this expectation requires to get in touch with the
ether-drift experiments (whose general aspects will be reviewed in
Sect.3) that indeed observe an instantaneous signal of this order of
magnitude but have interpreted so far this effect as spurious
instrumental noise.

Yet, some arguments might induce to modify this present
interpretation, at least in this particular context where one is
taking seriously the idea of an underlying superfluid ether. In
fact, the traditional analysis of these experiments is based on a
theoretical model where the hypothetical, preferred reference frame
is assumed to occupy a definite, fixed location in space. However,
suppose that the  superfluid ether exhibits a turbulent behaviour.
On the one hand, this poses the theoretical problem of how to relate
the macroscopic motions of the Earth's laboratory (daily rotation,
annual orbital revolution,...) to the microscopic measurement of the
speed of light inside the optical cavities. On the other hand, from
an experimental point of view, it suggests sizeable random
fluctuations of the signal that could be erroneously interpreted as
instrumental noise.

Since physical implications could be substantial, we believe that it
could be worth to perform some alternative test to check the
validity of the present interpretation. After all, other notable
examples are known (e.g. the CMBR) where, at the beginning, an
important physical signal was interpreted as a mere instrumental
effect. These more technical aspects of our analysis will be
discussed in Sects.4 and 5. Finally, Sect.6 will contain a summary
and our conclusions.

\section{Vacuum refractive index and effective acoustic metric}

As anticipated, the emergent-gravity approach derives from the
interesting analogies that one can establish between Einstein
gravity and the hydrodynamic limit of many physical systems in flat
space. However, to compare with experiments in weak gravitational
field, one should concentrate on the observed form of metric
structure and this restricts somehow the admissible types of
theoretical models. Thus, in order to set a definite framework for
our analysis, we shall consider the scenario sketched in Sect.1
where curvature arises due to modifications of the basic space-time
units. This general picture can be included in the emergent-gravity
philosophy provided one adopts some dynamical description where
these apparent curvature effects show up for length scales that are
much larger than any elementary particle, nuclear or atomic size
(e.g. a fraction of millimeter or so as in Ref.\cite{ultraweak}).
While this approach leads naturally to look for a non-zero light
anisotropy, a similar effect might also be expected if other
mechanisms are used to generate an effective geometry of the
acoustic form.

This idea of curvature as due to modifications of the basic
space-time units has been considered by several authors
\cite{wilson, dicke, puthoff} over the years and requires to first
adopt a "Lorentzian perspective" \cite{sonego} where physical rods
and clocks are held together by the same basic forces underlying the
structure of the 'ether' (the physical vacuum). Thus the principle
of relativity means that the measuring devices of moving observers
are dynamically affected in such a way that their uniform motions
become undetectable. In this representation, a gravitational field
is interpreted as a local modification in the state of the ether
such that now both the space-time units and the speed of light
become coordinate-dependent quantities thus generating an effective
curvature \cite{feybook,dicke1,atkinson}.

Let us now consider the problem of measuring the speed of light. On
a very general ground, to determine speed as (distance moved)/(time
taken), one must first choose some standards of distance and time
and different choices can give different answers. This is already
true in Special Relativity where the universal isotropic value $c$
entering Lorentz transformations is only obtained when describing
light propagation in an {\it inertial} frame. However, inertial
frames are just an idealization. Therefore, the appropriate
realization of this idea is to assume local standards of distance
and time such that the speed of light $c_\gamma$ is $c$ when
measured in a {\it freely falling} reference frame (at least in a
space-time region small enough that tidal effects can be neglected).
This provides the operative definition of the basic parameter $c$
entering Lorentz transformations.

With these premises, to describe light propagation in a vacuum
optical cavity, from the point of view of an observer $S'$ sitting
on the Earth's surface, one can adopt increasing degrees of
approximations:

~~~i) $S'$ is considered a freely-falling frame. Here, one starts
from the observation that the free-fall condition (in the
gravitational field of the Sun, of the other planets, of the
Galaxy,...) represents, up to tidal effects of the external
gravitational potential $U_{\rm ext}(x)$, the best approximation to
an inertial frame. In this first approximation one assumes
$c_\gamma=c$ so that, given two events which, in terms of the local
space-time units of the freely-falling observer, differ by $(dx, dy,
dz, dt)$, light propagation is described by the condition
(ff='free-fall') \BE\label{zero1} (ds^2)_{\rm ff}=c^2dt^2-
(dx^2+dy^2+dz^2)=0~\EE

~~ii) To a closer look, however, an observer placed on the Earth's
surface can only be considered a freely-falling frame up to the
presence of the Earth's gravitational field. Its inclusion leads to
tiny deviations from Eq.(\ref{zero1}). These can be estimated by
considering again $S'$ as a freely-falling frame, in the same
external gravitational field described by $U_{\rm ext}(x)$, that
however is also carrying on board a heavy object of mass $M$ (the
Earth's mass itself) that affects the effective local space-time
structure. To derive the required correction, let us again denote by
($dx$, $dy$, $dz$, $dt$) the local space-time units of the
freely-falling observer $S'$ in the limit $M=0$ and by $\delta U$
the dimensionless extra Newtonian potential produced by the heavy
mass $M$ at the experimental set up where one wants to describe
light propagation. Light propagation for the $S'$ observer can then
be described by the condition
\cite{pagano,ultraweak}\BE\label{iso}(ds^2)_{\rm \delta U}
={{c^2d\hat{t} ^2}\over{{\cal N}^2 }}-
(d\hat{x}^2+d\hat{y}^2+d\hat{z}^2)=0~\EE where, to first order in
$\delta U$, the space-time units ($d\hat{x}$, $d\hat{y}$,
$d\hat{z}$, $d\hat{t}$) are related to the corresponding ones ($dx$,
$dy$, $dz$, $dt$) for $\delta U=0$ through an overall re-scaling
factor \BE \label{lambda} \lambda= 1+|\delta U| \EE and
\BE\label{refractive1}{\cal N}= 1+2|\delta U|>1 \EE Therefore, to
this order, light is formally described as in General Relativity
where one finds the weak-field, isotropic form of the metric
\BE\label{gr} (ds^2)_{\rm GR}=c^2dT^2(1-2|U_{\rm N}|)-
(dX^2+dY^2+dZ^2)(1+2|U_{\rm N}|)\equiv c^2 d\tau^2 - dl^2\EE In
Eq.(\ref{gr}) $U_N$ denotes the Newtonian potential and ($dT$, $dX$,
$dY$, $dZ$) arbitrary coordinates defined for $U_{\rm N}=0$.
Finally, $d\tau$ and $dl$ denote the elements of proper time and
proper length in terms of which, in General Relativity, one would
again deduce from $ds^2=0$ the same universal value
$c={{dl}\over{d\tau}}$. This is the basic difference with
Eqs.(\ref{iso})-(\ref{refractive1}) where the physical unit of
length is $\sqrt {d\hat{x}^2+d\hat{y}^2+d\hat{z}^2}$, the physical
unit of time is $d\hat{t}$ and  instead a non-trivial refractive
index ${\cal N}$ is introduced. For an observer placed on the
Earth's surface,
 its value is \BE \label{refractive}{\cal N}-
1 \sim {{2G_N M}\over{c^2R}} \sim 1.4\cdot 10^{-9}\EE $G_N$ being
Newton's constant and $M$ and $R$ the Earth's mass and radius.

~~iii) Differently from General Relativity, in an interpretation
where ${\cal N}\neq 1$, the speed of light in the vacuum no longer
coincides with the parameter $c$ entering Lorentz transformations.
Therefore, as a general consequence of Lorentz transformations, an
isotropic propagation as in Eq.(\ref{iso}) can only be valid if the
Earth were at rest in a preferred frame $\Sigma$. In any other case,
one expects a non-zero anisotropy. To derive its value, one can
start from the original derivation of Jauch and Watson \cite{jauch}
who worked out the quantization of the electromagnetic field in a
moving medium of refractive index ${\cal N}$. They noticed that the
procedure introduces unavoidably a preferred frame, the one where
the photon energy does not depend on the direction of propagation,
and which is ``usually taken as the system for which the medium is
at rest". However, such an identification reflects the point of view
of Special Relativity with no preferred frame. More generally one
can adapt their results to the case where the angle-independence of
the photon energy defines some preferred frame $\Sigma$. Then, for
any non-zero velocity ${\bf V}$ of the Earth's laboratory, the mass
shell condition for the photon energy-momentum 4-vector $p_\mu\equiv
(E/c,{\bf p})$ \BE \label{masshell}
       p_\mu p_\nu g^{\mu\nu}= 0 ,
\EE is governed by the effective acoustic metric \BE g^{\mu\nu}=
\eta^{\mu\nu} + \kappa u^\mu u^\nu \EE where
 \BE \label{kappagamma}
       \kappa={\cal N}^2 - 1 \EE In the above relations, $\eta^{\mu\nu}$
indicates the Minkowski tensor and $u^\mu$ the dimensionless Earth's
velocity 4-vector $u^\mu\equiv(u^0,{\bf V}/c)$ with $u_\mu u^\mu=1$.
In coordinate space, the analogous condition is \BE
ds^2=g_{\mu\nu}dx^\mu dx^\nu=0 \EE with \BE
\label{effmetric}g_{\mu\nu}= \eta_{\mu\nu} -{{ {\cal N}^2 -
1}\over{{\cal N}^2 }}~ u_\mu u_\nu \EE and we have used the
relations $g^{\mu\lambda}g_{\lambda\nu}=\delta^\mu_\nu$ and
$u_\mu=\eta_{\mu\nu}u^\nu\equiv(u^0,-{\bf V}/c)$.

To first order in both $({\cal N}- 1)$ and $V/c$, the off-diagonal
elements \BE g_{0i}\sim 2({\cal N}- 1){{V_i}\over{c}} \EE can be
imagined as being due to a directional polarization of the vacuum
induced by the now moving Earth's gravitational field and express
the general property \cite{volkov} that any metric, locally, can
always be brought into diagonal form by suitable rotations and
boosts.

Quantitatively, Eq.(\ref{masshell}) gives a photon energy ($u^2_0=1
+ {\bf V}^2/c^2$) \BE \label{watson}
       E(| {\bf{p}}| , \theta)= c~{{ -\kappa u_0 \zeta
       + \sqrt{ |{\bf{p}}|^2(1+\kappa u^2_0) -
       \kappa \zeta^2 }}\over{ 1 + \kappa u^2_0}}
\EE with
\BE
       \zeta={\bf{p}}\cdot{{{\bf{V}}}\over{c}}= |{\bf{p}}|\beta \cos\theta ,
\EE where $\beta={{|{\bf{V}}|}\over{c}}$ and
$\theta\equiv\theta_{\rm lab}$ indicates the angle defined, in the
laboratory $S'$ frame, between the photon momentum and ${\bf{V}}$.
By using the above relation, one gets the one-way speed of light in
the $S'$ frame
\begin{eqnarray}
       & &{{E(| {\bf{p}}| , \theta)}\over{|{\bf{p}}|}}=
       c_\gamma(\theta)= c~{{ -\kappa \beta  \sqrt{1+\beta ^2} \cos\theta
+ \sqrt{ 1+ \kappa+ \kappa \beta^2 \sin^2\theta} }
       \over{1+ \kappa(1+\beta^2)}} .
\end{eqnarray}
or to ${\cal O}(\kappa)$ and ${\cal O}(\beta^2)$
\BE \label{oneway}
       c_\gamma(\theta)= {{c} \over{{\cal N}}}~\left[
       1- \kappa \beta \cos\theta -
       {{\kappa}\over{2}} \beta^2(1+\cos^2\theta)\right]
\EE
Further, one can compute the two-way speed
\begin{eqnarray}
\label{twoway}
       \bar{c}_\gamma(\theta)&=&
       {{ 2  c_\gamma(\theta) c_\gamma(\pi + \theta) }\over{
       c_\gamma(\theta) + c_\gamma(\pi + \theta) }} \nonumber \\
       &\sim& {{c} \over{{\cal N}}}~\left[1-\beta^2\left(\kappa -
       {{\kappa}\over{2}} \sin^2\theta\right) \right]
\end{eqnarray}
and define the RMS \cite{rms} anisotropy parameter $B$ through the
relation \footnote{There is a subtle difference between our
Eqs.(\ref{oneway}) and(\ref{twoway}) and the corresponding Eqs. (6)
and (10) of Ref.~\cite{pla} that has to do with the relativistic
aberration of the angles. Namely, in Ref.\cite{pla}, with the
(wrong) motivation that the anisotropy is ${\cal O}(\beta^2)$, no
attention was paid to the precise definition of the angle between
the Earth's velocity and the direction of the photon momentum. Thus
the two-way speed of light in the $S'$ frame was parameterized in
terms of the angle $\theta\equiv\theta_\Sigma$ as seen in the
$\Sigma$ frame. This can be explicitly checked by replacing in our
Eqs.~(\ref{oneway}) and(\ref{twoway}) the aberration relation
$\cos \theta_{\rm lab}=(-\beta + \cos\theta_\Sigma)/
       (1-\beta\cos\theta_\Sigma)$
or equivalently by replacing $\cos \theta_{\Sigma}=(\beta +
\cos\theta_{\rm lab})/ (1+\beta\cos\theta_{\rm lab})$ in Eqs. (6)
and (10) of Ref.~\cite{pla}. However, the apparatus is at rest in
the laboratory frame, so that the correct orthogonality condition of
two optical cavities at angles $\theta$ and $\pi/2 + \theta$ is
expressed in terms of $\theta=\theta_{\rm lab}$ and not in terms of
$\theta=\theta_{\Sigma}$. This trivial remark produces however a
non-trivial difference in the value of the anisotropy parameter. In
fact, the correct resulting $|B|$ Eq. (\ref{rmsb}) is now smaller by
a factor of 3 than the one computed in Ref.\cite{pla} by adopting
the wrong definition of orthogonality in terms of
$\theta=\theta_{\Sigma}$.}
\BE \label{rms}
       {{\bar{c}_\gamma(\pi/2 +\theta)- \bar{c}_\gamma (\theta)} \over
       {\langle \bar{c}_\gamma \rangle }} \sim
       B{{V^2 }\over{c^2}} \cos(2\theta) \EE
with
\BE \label{rmsb}
       |B|\sim {{\kappa}\over{2}}\sim {\cal N}-1
\EE
From the previous analysis, by replacing the value of the refractive
index Eq.(\ref{refractive}) and adopting, as a rough order of
magnitude, the typical value of most cosmic motions $V\sim$ 300
km/s, one expects an average fractional anisotropy \BE
\label{averani}
        {{\langle\Delta \bar{c}_\theta \rangle} \over{c}} \sim
       |B|{{V^2 }\over{c^2}} ={\cal O}(10^{-15}) \EE
that could finally be detected in ether-drift experiments by
measuring the beat frequency $\Delta \nu$ of two orthogonal
cavity-stabilized lasers.

\section{Ether-drift experiments in a superfluid vacuum}

In ether-drift experiments, the search for time modulations of the
signal that might be induced by the Earth's rotation (and its
orbital revolution) has always represented a crucial ingredient for
the analysis of the data. For instance, let us consider the relative
frequency shift of two optical resonators for the experiment of
Ref.\cite{peters} \BE \label{basic2}
      {{\Delta \nu (t)}\over{\nu_0}} =
      {S}(t)\sin 2\omega_{\rm rot}t +
      {C}(t)\cos 2\omega_{\rm rot}t
\EE
where $\omega_{\rm rot}$ is the rotation frequency of one resonator
with respect to the other which is kept fixed in the laboratory and
oriented north-south. If one assumes that, for short-time
observations of 1-2 days, the time dependence of a hypothetical
physical signal can only be due to (the variations of the projection
of ${\bf V}$ in the interferometer's plane caused by) the Earth's
rotation, $S(t)$ and $C(t)$ admit the simplest Fourier expansion
($\tau=\omega_{\rm sid}t$ is the sidereal time of the observation in
degrees) \cite{peters} \BE \label{amorse1}
      {S}(t) = S_0 +
      {S}_{s1}\sin\tau +{S}_{c1} \cos\tau
       + {S}_{s2}\sin(2\tau) +{S}_{c2} \cos(2\tau)
\EE \BE \label{amorse2}
      {C}(t) = {C}_0 +
      {C}_{s1}\sin\tau +{C}_{c1} \cos\tau
       + {C}_{s2}\sin(2\tau) +{C}_{c2} \cos(2\tau)
\EE
with time-independent $C_k$ and $S_k$ Fourier coefficients.
Therefore, by accepting this theoretical framework, it becomes
natural to average the various $C_k$ and $S_k$ over any 1-2 day
observation period. By further averaging over many short-period
experimental sessions, the general conclusion \cite{joint,newberlin}
is that, although the typical instantaneous signal is ${\cal
O}(10^{-15})$, the global averages $(C_k)^{\rm avg}$ and $(S_k)^{\rm
avg}$  for the Fourier coefficients are at the level ${\cal
O}(10^{-17})$ or smaller and, with them, the derived parameters
entering the SME \cite{sme} and RMS \cite{rms} models.

However, by taking seriously a flat-space origin of curvature from
the distortions of an underlying, superfluid quantum ether, there
might be different types of ether-drift where this straightforward
averaging procedure is {\it not} allowed. Then, the same basic
experimental data might admit a different interpretation and a
definite instantaneous signal $\Delta \nu (t)\neq 0$ could become
consistent with $(C_k)^{\rm avg} \sim (S_k)^{\rm avg}\sim 0$.

For this reason, we believe that, by accepting the idea that there
might be a preferred reference frame, which is the modern
denomination of the old ether, before assuming any definite
theoretical scenario, one should first ask: if light were really
propagating in a physical medium, an ether, and not in a trivial
empty vacuum, how should the motion of (or in) this medium be
described ? Namely, could this relative motion exhibit variations
that are {\it not} only due to known effects as the Earth's rotation
and orbital revolution ?

Without fully understanding the nature of that substratum that we
call physical vacuum, it is not possible to make definite
predictions. Still, according to present elementary-particle theory,
this physical vacuum is not trivially empty but is filled by
particle condensates \cite{mech,volo} and therefore it becomes
natural to represent the vacuum as a superfluid medium, a quantum
liquid. By further considering the idea of a non-zero vacuum energy,
this physical substratum could also represent a preferred reference
frame \cite{vacuum}. In this picture, the standard assumption of
smooth sinusoidal variations of the signal, associated with the
Earth's rotation and its orbital revolution, corresponds to describe
the superfluid flow in terms of simple regular motions.

However, visualization techniques that record the flow of superfluid
helium show \cite{zhang} the formation of turbulent structures with
a velocity field that fluctuates randomly around some average value.
In our case, the concept of turbulence arises naturally if one takes
seriously the idea of the vacuum as a quantum liquid, i.e. a fluid
where density and current satisfy local uncertainty relations, as
suggested by Landau \cite{landausuper} to explain the phenomenon of
superfluidity. According to this quantum-hydrodynamical
representation, a fluid whose density is exactly known at some point
becomes, at that same point, totally undetermined in its velocity.
Therefore a (nearly) incompressible quantum liquid should be thought
as microscopically turbulent.

While this provides some motivation to look for an ultimate quantum
origin of turbulence \cite{yogi} and for the striking similarities
\cite{vinen} between many aspects of turbulence in fluids and
superfluids, this picture of the vacuum allows to establish a link
with the old perspective where the ether was providing the support
for the electromagnetic waves. In fact, it is known
\cite{troshkin,tsankov,marmanis} that one can establish a formal
equivalence  between the propagation of small disturbances in an
incompressible turbulent fluid and the propagation of
electromagnetic waves as described by Maxwell equations. To this
end, one has to decompose all basic quantities of the fluid
(pressure, velocity, density) into an average background and
fluctuating components and then linearize the hydrodynamical
equations. By using this method, Puthoff \cite{puthoff2} has been
able to extend the analogy to general relativity by deriving
effective metric coefficients of the type needed to account for
`gravitomagnetic' and `gravitoelectric' effects. For all these
reasons, the idea of a turbulent quantum ether becomes a natural
representation of the physical vacuum.

To exploit the possible implications for ether-drift experiments,
let us first recall the general aspects of any turbulent flow. This
is characterized by extremely irregular variations of the velocity,
with time at each point and between different points at the same
instant, due to the formation of eddies \cite{landau}. For this
reason, the velocity continually fluctuates about some mean value
and the amplitude of these variations is {\it not} small in
comparison with the mean velocity itself. The time dependence of a
typical turbulent velocity field can be expressed as \cite{landau}
\BE {\bf v}(x,y,z,t)=\sum_{p_1p_2..p_n} {\bf
a}_{p_1p_2..p_n}(x,y,z)\exp(-i\sum^{n}_{j=1}p_j\phi_j) \EE where the
quantities $\phi_j=\omega_j t+ \beta_j $ vary with time according to
fundamental frequencies $\omega_j$ and depend on some initial phases
$\beta_j$. As the Reynolds number ${\cal R}$ increases, the total
number $n$ of $\omega_j$ and $\beta_j$ increases. In the ${\cal R}
\to \infty$ limit, their number diverges so that the theory of such
a turbulent flow must be a statistical theory.

Now, due to the presumably vanishingly small viscosity of a
superfluid ether, the relevant Reynolds numbers are likely
infinitely large in most regimes and we might be faced precisely
with such limit of the theory where the temporal analysis of the
flow requires an infinite number of frequencies and the physical
vacuum behaves as a {\it stochastic} medium. In this case random
fluctuations of the signal, superposed on the smooth sinusoidal
behaviour associated with the Earth's rotation (and orbital
revolution), would produce deviations of the time dependent
functions $S(t)$ and $C(t)$ from the simple structure in
Eqs.(\ref{amorse1}) and (\ref{amorse2}) and an effective temporal
dependence of the fitted $C_k=C_k(t)$ and $S_k=S_k(t)$. In this
situation, due to the strong cancelations occurring in vectorial
quantities when dealing with stochastic signals, one could easily
get vanishing global inter-session averages \BE (C_k)^{\rm avg} \sim
(S_k)^{\rm avg} \sim 0\EE Nevertheless, as it happens with the
phenomena affected by random fluctuations, the average quadratic
amplitude of the signal could still be preserved. Namely, by
defining the positive-definite amplitude $A(t)$ of the signal \BE
{{\Delta \nu (t)}\over{\nu_0}}= A(t)e^{i\Phi(t)} \EE where \BE A(t)=
\sqrt{S^2(t) +C^2(t)} \EE a definite non-zero $\langle A\rangle$
might well coexist with $(C_k)^{\rm avg} \sim (S_k)^{\rm avg}\sim
0$. Physical conclusions would then require to compare the obtained
value of $\langle A\rangle$ with the short-term, stability limits of
the individual resonators.

\section{Noise or stochastic turbulence ?}

To provide some evidence that indeed, in ether-drift experiments, we
might be faced with stochastic fluctuations of a physical signal, we
have first considered the experimental apparatus of
Ref.\cite{crossed} where, to minimize all sources of systematic
asymmetry, the two optical cavities were obtained from the same
monolithic block of ULE (Ultra Low Expansion material). In these
conditions, due to sophisticated electronics and temperature
controls, the short-term (about 40 seconds) stability limits for the
individual optical cavities are extremely high. Namely, for the
non-rotating set up, by taking into account all possible systematic
effects, one deduces a stability of better than $\pm 0.05$ Hz for
the individual cavities and thus better than $\pm 2\cdot 10^{-16}$
in units of a laser frequency $\nu_0=2.82\cdot 10^{14}$ Hz. This is
of the same order of the {\it average} frequency shift between the
two resonators, say $(\Delta \nu)^{\rm avg} \lesssim \pm 0.06$ Hz,
when averaging the signal over a very large number of temporal
sequences (see their Fig.9b).

However, the magnitude of the {\it instantaneous} frequency shift is
much larger, say $\pm 1$ Hz (see their Fig.9a), and so far has been
interpreted as spurious instrumental noise. To check this
interpretation, we observe that, in the absence of any light
anisotropy, the noise in the beat frequency should be comparable to
the noise of the individual resonators. Instead, for the same
non-rotating set up, the minimum noise in the beat signal was found
to be 10 times bigger, namely $1.9\cdot 10^{-15}$ (see Fig.8 of
Ref.\cite{crossed}). Also the trend of the noise in the beat signal,
as function of the averaging time, is different from the
corresponding one observed in the individual resonators thus
suggesting that the two types of noise might have different origin.

The authors tend to interpret this relatively large beat signal as
cavity thermal noise and refer to \cite{numata}. However, this
interpretation is not so obvious since the same noise in the {\it
individual} cavities was reduced to a much lower level.

Similar conclusions can be obtained from the more recent analysis of
Ref.\cite{newberlin} where the stability of the individual
resonators is at the same level $10 ^{-16}$. Nevertheless, the
typical $C(t)$ and $S(t)$ entering the beat signal are found in the
range $ \pm 10^{-15}$ (see their Fig.4a) and are again interpreted
in terms of cavity thermal noise.

In any case, as an additional check, one can always compare with
other experiments performed in the cryogenic regime. If this typical
$ {\cal O}(10^{-15})$ beat signal reflects the stochastic nature of
an underlying quantum ether (and is not just an instrumental
artifact of the resonating cavities) it should be found in these
different experiments as well.

\section{An alternative analysis of the data}

Motivated by the previous arguments, we have decided to explore the
idea that the observed beat signal between vacuum optical resonators
could be due to some form of turbulent ether flow. This poses the
problem to relate the macroscopic motions of the Earth's laboratory
(daily rotation, annual orbital revolution,...) to the microscopic
nature of the measurement of the speed of light inside the optical
cavities. For very large Reynolds numbers, some macroscopic
directional effects can be lost in the reduction process as energy
is transferred to smaller and smaller scales. Thus, even though the
relevant Earth's cosmic motion corresponds to that indicated by the
anisotropy of the CMBR ($V\sim $370 km/s, angular declination
$\gamma\sim -6$ degrees, and right ascension $\alpha \sim$ 168
degrees) it might be difficult to detect these parameters in the
laboratory.

In this perspective, one should abandon the previous type of
analysis based on assuming a fixed preferred reference frame and
extract the amplitude $A(t)$ of the signal from the instantaneous
data obtained from a few rotations of the interferometer {\it
before} any averaging procedure. As anticipated, by inspection of
Fig.4a of Ref.\cite{newberlin} the typical $C(t)$ and $S(t)$
entering the beat signal are found in the range $\pm
12\cdot10^{-16}$ and this fixes the typical size of $A(t)$. However,
the instantaneous values cannot be extracted from the figure.
Therefore, in this condition, to obtain a rough estimate, we shall
try to evaluate $\langle A\rangle$ from the $C_k$ and $S_k$ Fourier
coefficients obtained after averaging the signal within each
short-period session.

For our analysis, we have first re-written Eq.(\ref{basic2}) as \BE
\label{basic3}
      {{\Delta \nu (t)}\over{\nu_0}} =
      A(t)\cos (2\omega_{\rm rot}t -2\theta_0(t))
\EE with \BE \label{interms}
C(t)=A(t)\cos2\theta_0(t)~~~~~~~~S(t)=A(t)\sin2\theta_0(t)\EE
$\theta_0(t)$ representing the instantaneous direction of the
ether-drift effect in the plane of the interferometer. In this
plane, the projection of the full ${\bf V}$ is specified by its
magnitude $v=v(t)$ and by its direction $\theta_0=\theta_0(t)$
(counted by convention from North through East so that North is
$\theta_0=0$ and East is $\theta_0=\pi/2$). If one assumes
Eqs.(\ref{amorse1}) and (\ref{amorse2}), then $v(t)$ and
$\theta_0(t)$ can be obtained from the relations
\cite{nassau,dedicated} \BE \label{cosine}
       \cos z(t)= \sin\gamma\sin \phi + \cos\gamma
       \cos\phi \cos(\tau-\alpha)
\EE \BE
       \sin z(t)\cos\theta_0(t)= \sin\gamma\cos \phi -\cos\gamma
       \sin\phi \cos(\tau-\alpha)
\EE \BE
       \sin z(t)\sin\theta_0(t)= \cos\gamma\sin(\tau-\alpha) \EE
\BE \label{projection}
       v(t)=V \sin z(t) ,
\EE where $\alpha$ and $\gamma$ are respectively the right ascension
and angular declination of ${\bf{V}}$. Further, $\phi$ is the
latitude of the laboratory and $z=z(t)$ is the zenithal distance of
${\bf{V}}$. Namely, $z=0$ corresponds to a ${\bf{V}}$ which is
perpendicular to the plane of the interferometer and $z=\pi/2$ to a
${\bf{V}}$ that lies entirely in that plane. From the above
relations, by using the ${\cal O}(v^2/c^2)$ relation $A(t) \sim
{{v^2(t)}\over{c^2}}$, the other two amplitudes $S(t)=A(t)\sin
2\theta_0(t)$ and $C(t)=A(t)\cos 2\theta_0(t)$ can be obtained up to
an overall proportionality constant. By using the expressions for
$S(t)$ and $C(t)$ reported in Table I of Ref.~\cite{peters} (in the
RMS formalism \cite{rms}), this proportionality constant turns out
to be ${{1}\over{2}}|B|$  so that we finally find the basic relation
\BE \label{amplitude1}
       A(t)= {{1}\over{2}}|B| {{v^2(t) }\over{c^2}}
\EE where $B$ is the anisotropy parameter entering the two-way speed
of light Eq.(\ref{rms}). It is a simple exercise to check that, by
using Eqs.(\ref{interms}), Eqs.(\ref{cosine})-(\ref{amplitude1}) and
finally replacing $\chi=90^o-\phi$, one re-obtains the expansions
for $C(t)$ and $S(t)$ reported in Table I of Ref.~\cite{peters}.

We can then replace Eq.~(\ref{projection}) into
Eq.~(\ref{amplitude1}) and, by adopting a notation of the type in
Eqs.(\ref{amorse1})-(\ref{amorse2}), express the Fourier expansion
of $A(t)$ as
\BE \label{amorse}
       A(t) = A_0 +
       A_1\sin\tau +A_2 \cos\tau
        +  A_3\sin(2\tau) +A_4 \cos(2\tau)
\EE
where
\BE \label{aa0}
   \langle
A\rangle  =   A_0 ={{1}\over{2}}|B| {{\langle
v^2(t)\rangle}\over{c^2}}=
       {{1}\over{2}}|B| {{V^2}\over{c^2}}
       \left(1- \sin^2\gamma\cos^2\chi
       - {{1}\over{2}} \cos^2\gamma\sin^2\chi \right)
\EE
\BE \label{a1}
       A_1=-{{1}\over{4}}|B| {{V^2}\over{c^2}}\sin 2\gamma
       \sin\alpha \sin 2\chi
~~~~~~~~~~~~~~~
       A_2=-{{1}\over{4}}|B| {{V^2}\over{c^2}}\sin 2\gamma
       \cos\alpha \sin 2\chi
\EE
\BE \label{a3}
       A_3=-{{1}\over{4}}  |B| {{V^2}\over{c^2}}\cos^2 \gamma
       \sin 2\alpha \sin^2 \chi
~~~~~~~~~~~~~~~
       A_4=-{{1}\over{4}} |B| {{V^2}\over{c^2}} \cos^2 \gamma
       \cos 2\alpha \sin^2 \chi\EE
and we have denoted by $\langle..\rangle$ the daily average of a
quantity (not to be confused with the inter-session experimental
averages denoted by $(..)^{\rm avg}$).

To obtain $A_0$ from the $C_k$ and $S_k$, we observe that by using
Eq.(\ref{amorse}) one obtains \BE \label{amplitude0}\langle ~A^2(t)~
\rangle= A^2_0+ {{1}\over{2}}(A^2_{1}+A^2_{2}+A^2_{3}+A^2_{4})\EE On
the other hand, by using Eqs.(\ref{amorse1}), (\ref{amorse2}) and
(\ref{interms}), one also obtains
 \BE
\label{amplitude}\langle ~A^2(t)~ \rangle= \langle ~C^2(t) +S^2(t)~
\rangle=C^2_0 + S^2_0 + Q^2 \EE where \BE \label{Q} Q= \sqrt{
{{1}\over{2}}(C^2_{11}+S^2_{11}+C^2_{22}+S^2_{22})} \EE and  \BE
\label{csid}
      {C}_{11}\equiv \sqrt{{C}^2_{s1}
      + {C}^2_{c1}}
~~~~~~~~~~~~~~~~
      {C}_{22}\equiv \sqrt{{C}^2_{s2}
      + {C}^2_{c2}}
\EE
 \BE \label{s2sid}
      {S}_{11}\equiv \sqrt{{S}^2_{s1}
      + {S}^2_{c1}}
~~~~~~~~~~~~~~~~
 {S}_{22}\equiv \sqrt{{S}^2_{s2}
      + {S}^2_{c2}}
\EE
Therefore, one can combine the two relations and get \BE
\label{final} A^2_0(1+r)= C^2_0 + S^2_0 + Q^2 \EE where
 \BE r\equiv
{{1}\over{2A^2_0}}(A^2_{1}+A^2_{2}+A^2_{3}+A^2_{4}) \EE By computing
the ratio $r=r(\gamma,\chi)$ with Eqs.(\ref{aa0})-(\ref{a3}), one
finds \BE \label{range} 0\leq r\leq 0.4\EE for the latitude of the
laboratories in Berlin \cite{peters} and D\"usseldorf
\cite{schiller} in the full range $0 \leq |\gamma|\leq \pi/2$.  We
can thus define an average amplitude, say $\hat{A}_0$, which is
determined in terms of $Q$ alone as \BE \label{AQ} \hat{A}_0 \equiv
{{Q}\over{\sqrt{1+r}}} \sim (0.92 \pm 0.08)Q \EE where the
uncertainty takes into account the numerical range of $r$ in
Eq.(\ref{range}). This quantity provides, in any case, a {\it lower
bound} for the true experimental $\langle A\rangle$ since \BE
\label{lower} \langle A \rangle=A_0=\sqrt{ {{C^2_0 + S^2_0
+Q^2}\over{1+r}}
 } \geq {{Q}\over{\sqrt{1+r}}}\equiv \hat{A}_0 \EE
At the same time $Q$ is determined only by the $C_{s1}$,
$C_{c1}$,... and their S-counterparts. According to the authors of
Refs.\cite{peters,joint}, these coefficients are much less affected
by spurious effects, as compared to $C_0$ and $S_0$, and so will be
our amplitude $\hat{A}_0$.

\begin{table}
\caption{We report the various values of $Q$, their uncertainties
$\Delta Q$ and the ratio $R=Q/\Delta Q$ for each of the 27
experimental sessions of Ref.\cite{joint}. These values have been
extracted, by using Eqs. (\ref{Q}), (\ref{csid}) and (\ref{s2sid})
and according to standard error propagation for a composite
observable, from the basic $C_k$ and $S_k\equiv B_k$ coefficients
reported in Fig.2 of Ref.\cite{joint}.}
\begin{center}
\label{tab:3}
\begin{tabular}{lll}
\hline\noalign{\smallskip}
 $ Q  [{\rm x}10^{-16}]$ & $ \Delta Q  [{\rm
x}10^{-16}]$ & $R= Q/\Delta Q  $  \\
\noalign{\smallskip}\hline\noalign{\smallskip} \hline
$13.3$ & ~~~ $3.4$ &~~~  $3.9$  \\
$14.6$ & ~~~ $4.8$ & ~~~ $3.0$  \\
$6.6$ &  ~~~ $2.6$ & ~~~ $2.5$  \\
$17.8$ & ~~~ $2.8$ & ~~~ $6.3$ \\
$14.0$ & ~~~ $5.8$ & ~~~ $2.5$ \\
$11.1$ & ~~~ $4.2$ & ~~~ $2.6$ \\
$13.0$ & ~~~ $4.2$ & ~~~ $3.1$ \\
$19.2$ & ~~~ $6.1$ & ~~~ $3.1$ \\
$13.0$ & ~~~ $4.7$ & ~~~ $2.8$ \\
$12.0$ & ~~~ $3.5$ & ~~~ $3.4$ \\
$5.7$ &  ~~~ $2.4$ &  ~~~ $2.4$ \\
$14.6$ &  ~~~ $5.2$ & ~~~ $2.8$ \\
$16.9$ &  ~~~ $3.3$ & ~~~ $5.1$ \\
$8.3$ &  ~~~ $2.4$ &  ~~~ $3.4$ \\
$27.7$ & ~~~ $4.5$ & ~~~ $6.2$  \\
$28.3$ & ~~~ $5.7$ & ~~~ $5.0$  \\
$12.7$ & ~~~ $2.5$ & ~~~ $5.1$ \\
$12.1$ & ~~~ $5.3$ & ~~~ $2.3$ \\
$13.7$ & ~~~ $6.0$ & ~~~ $2.3$ \\
$23.9$ & ~~~ $5.7$ & ~~~ $4.2$ \\
$28.9$ & ~~~ $4.3$ & ~~~ $6.7$ \\
$18.4$ & ~~~ $5.1$ & ~~~ $3.6$ \\
$19.2$ & ~~~ $6.2$ & ~~~ $3.1$ \\
$11.9$ & ~~~ $2.7$ & ~~~ $4.4$ \\
$18.1$ & ~~~ $5.4$ & ~~~ $3.3$ \\
$4.2$ & ~~~  $2.9$ &  ~~~ $1.4$ \\
$31.6$ &~~~  $7.9$ & ~~~ $4.0$ \\
\noalign{\smallskip}\hline
\end{tabular}
\end{center}
\end{table}

By starting from the basic data for the $C_k$ and $S_k$ reported in
 in Fig.2 of Ref.\cite{joint} we have thus computed the $Q$ values for the
27 short-period experimental sessions. Their values are reported in
Table I. These data represent, within their statistics, a sufficient
basis to deduce that a rather stable pattern is obtained. This is
due to the rotational invariant character of $Q$ in the 8-th
dimensional space of the $C_{s1}$, $C_{c1}$,...$S_{s2}$, $S_{c2}$ so
that variations of the individual coefficients tend to compensate.
By taking an average of these 27 determinations one finds a mean
value \BE \label{Qmean1} (Q)^{\rm avg}=\left(13.0 \pm 0.7 \pm
3.8\right)\cdot 10^{-16} ~~~~~~~~~~~~{\rm Ref.\cite{joint}}\EE where
the former error is purely statistical and the latter represents an
estimate of the systematical effects.

As anticipated, for a further control of the validity of our
analysis, we have compared with the cryogenic experiment of
Ref.\cite{schiller}. In this case, we have obtained the analogous
value \BE Q=(13.1 \pm 2.1)\cdot 10^{-16}~~~~~~~~~~~~~~~{\rm
Ref.\cite{schiller}} \EE  from the corresponding $C_k$ and $S_k$
coefficients. Thus, by using Eq.(\ref{AQ}) and the two values of $Q$
reported above, we obtain \BE \label{final1} (\hat{A}_0)^{\rm
avg}=(12.0 \pm 1.0 \pm 3.5)\cdot 10^{-16}~~~~~~~~~~~~{\rm
Ref.\cite{joint}} \EE \BE \label{mean2} \hat{A}_0= (12.1 \pm 1.0 \pm
2.1)\cdot 10^{-16}~~~~~~~~~~~~~~~~~~{\rm Ref.\cite{schiller}}\EE
where the former uncertainty takes into account the variation of $r$
in Eq.(\ref{range}) and the latter is both statistical and
systematical.

We emphasize that this stable value of about $10^{-15}$ is unlike to
represent just a spurious instrumental artifact of the optical
cavities as that discussed in Ref.\cite{numata}. In fact, the
estimate of Ref.\cite{numata} is based on the
fluctuation-dissipation theorem, and therefore there is no real
reason that both room temperature and cryogenic experiments exhibit
the same experimental noise.

In conclusion, our alternative analysis confirms an average
experimental amplitude \BE \label{bexp3}\langle A \rangle _{\rm
exp}\sim 10^{-15} \EE that can be compared with our theoretical
prediction based on Eqs.(\ref{refractive}), (\ref{rmsb})and
(\ref{amplitude1}) \BE \label{theory3} \langle A \rangle _{\rm th} =
{{1}\over{2}}({\cal N}- 1){{\langle v^2(t)\rangle}\over{c^2}} \sim~
7\cdot 10^{-10}~ {{\langle v^2(t)\rangle}\over{c^2}}
 \EE
Therefore, by assuming the typical speed $\sqrt {\langle
v^2(t)\rangle}\sim$ 300 km/s of most cosmic motions, one predicts a
theoretical value \BE \label{theory4} \langle A \rangle _{\rm th}
\sim 7\cdot 10^{-16} \EE in good agreement with the experimental
result Eq.(\ref{bexp3}).

\section{Summary and conclusions}

In the framework of the so called emergent-gravity approach, the
space-time curvature observed in a gravitational field is
interpreted as an effective phenomenon originating from
long-wavelength fluctuations of the same physical, flat-space
vacuum, viewed as a form of superfluid quantum ether. Thus, this
view is similar to a hydrodynamic description of moving fluids where
curvature arises on length scales that are much larger than the size
of the elementary constituents of the fluid. This is the basic
difference with the more conventional point of view where, instead,
curvature represents a fundamental property of space-time that can
also show up at arbitrarily small length scales.

In principle, being faced with two different interpretations of the
same, observed metric structure, one might ask if this basic
conceptual difference could have phenomenological implications. We
have argued in Sects. 1 and 2 that, in an approach where an
effective curvature emerges from modifications of the basic
space-time units, it is a pure experimental issue whether the
velocity of light $c_\gamma$, which is measured inside a vacuum
optical cavity, coincides or not with the basic parameter $c$
entering Lorentz transformations. Thus, it makes sense to consider a
scenario where $c_\gamma \neq c$ and the idea of an angular
anisotropy of the two-way speed ${{\langle\Delta \bar{c}_\theta
\rangle}\over{c}}$. This could be detected through the frequency
shift \BE
      {{\Delta \nu (t)}\over{\nu_0}} =
      {S}(t)\sin 2\omega_{\rm rot}t +
      {C}(t)\cos 2\omega_{\rm rot}t
\EE of two rotating optical resonators in those ether-drift
experiments that represent the modern version of the original
Michelson-Morley experiment. As indicated at the end of Sect.2, for
an apparatus placed on the Earth's surface, this fractional
asymmetry is expected to be ${{\langle\Delta \bar{c}_\theta \rangle}
\over{c}}={\cal O}(10^{-15})$ and this should be compared with the
experimental data.

Now, the present experiments indeed observe an instantaneous signal
that has precisely this order of magnitude but have interpreted so
far this effect as spurious instrumental noise. The point is that
the traditional analysis of the data is based on a theoretical model
where the hypothetical preferred reference frame is assumed to
occupy a definite, fixed location in space. Thus a true physical
signal has always been searched through smooth, sinusoidal
modulations associated with the Earth's rotation (and its orbital
revolution).

However, we have also argued in Sect.3 that, in this particular
context, where one is taking seriously the idea of an underlying
superfluid quantum ether, one might also consider unconventional
forms of ether-drift and alternative interpretation of the
experimental data. For instance, some theoretical arguments suggest
that the superfluid ether might be in a turbulent state of motion
thus making non-trivial to relate the macroscopic motions of the
Earth's laboratory (daily rotation, annual orbital revolution,...)
to the microscopic measurement of the speed of light inside the
optical cavities. In this scenario, where the physical vacuum
behaves as a stochastic medium, a true physical signal might exhibit
sizeable random fluctuations that could be erroneously interpreted
as instrumental noise.

For this reason, by following the analysis of our Sects. 4 and 5, we
propose a consistency check of the present interpretation of the
data. This alternative analysis requires to first introduce the
instantaneous magnitude $v=v(t)$ and direction
$\theta_0=\theta_0(t)$ of the hypothetical ether-drift effect
(projected in the plane of the interferometer). In terms of these
two basic parameters, one can re-write the two amplitudes $C(t)$ and
$S(t)$ as \BE
C(t)=A(t)\cos2\theta_0(t)~~~~~~~~S(t)=A(t)\sin2\theta_0(t)\EE where
\BE
       A(t)= {{1}\over{2}}|B| {{v^2(t) }\over{c^2}}
\EE and $B$ is the anisotropy parameter entering the two-way speed
of light Eq.(\ref{rms}). By first concentrating on the amplitude \BE
A(t)= \sqrt{S^2(t) +C^2(t)}\EE (which is less dependent on the
fluctuating directional aspects of the signal) one should compare
its experimental value with the typical short-term stability of the
individual resonators. Since this individual stability, in today's
most precise experiments, is at the level $10^{-16}$ the actual
experimental value $\langle A(t)\rangle_{\rm exp} \sim 10^{-15}$ is
about ten times larger and might not be a spurious instrumental
effect. Moreover, this measured value is completely consistent with
the average theoretical expectation
Eqs.(\ref{theory3})-(\ref{theory4}), namely $\langle A(t) \rangle
_{\rm th} \sim 7\cdot 10^{-16}$, for the typical speed $\sqrt
{\langle v^2(t)\rangle}\sim$ 300 km/s of most cosmic motions.

Thus we look forward to a new analysis of the raw data, before any
averaging procedure, that one could start, for instance, by
considering the daily plots of $A(t)=\sqrt{S^2(t) +C^2(t)}$ in terms
of the 13384 individual determinations of $S(t)$ and $C(t)$ reported
in Fig.4a of Ref.\cite{newberlin} (that, in their present form,
cannot be used by the reader). In the end, from a new set of
precious, combined informations, the observed frequency shift,
rather than being spurious noise of the underlying optical cavities,
might turn out to reflect two basic properties of the physical
vacuum. Namely, this could be a polarizable medium responsible for
the apparent curvature effects seen in a gravitational field and, at
the same time, a stochastic medium, similar to a superfluid in a
turbulent state of motion, responsible for the observed strong
random fluctuations of the signal. All together, the situation might
resemble the discovery of the CMBR that, at the beginning, was also
interpreted as mere instrumental noise.

After this first series of checks, further tests could be performed
by placing the interferometer on board of a spacecraft, as in the
OPTIS proposal \cite{optis}. In this case where, even in a
flat-space picture, the vacuum refractive index ${\cal N}$ for the
freely-falling observer is exactly unity, the typical instantaneous
$\Delta \nu$ should be much smaller (by orders of magnitude) than
the corresponding  ${\cal O}(10^{-15})$ value measured with the same
interferometer on the Earth's surface. Such a substantial reduction
of the instantaneous signal, in a gravity-free environment, would be
extremely important for our understanding of gravity. \vskip 20 pt
\centerline{{\bf Acknowledgments}} We thank  E. Costanzo for
collaboration at the initial stage of this work.

\vfill\eject

\end{document}